\documentclass[journal]{IEEEtran}

\usepackage[dvipsnames]{xcolor}
\usepackage[acronym]{glossaries}
\usepackage[colorlinks, citecolor=blue, linkcolor=blue]{hyperref}
\usepackage[textwidth=1.6cm]{todonotes}
\usepackage{comment}
\usepackage{nameref}
\usepackage{subcaption}
\usepackage{multicol}
\usepackage{multirow}
\usepackage{rotating}
\usepackage{stfloats}
\usepackage[noadjust]{cite}
\usepackage{float}
\usepackage{amsfonts}
\usepackage{upgreek}
\usepackage{mathtools}
\usepackage{scalerel}
\usepackage{braket}
\usepackage{soul}
\usepackage{tcolorbox}

\newcommand{\widebar}[1]{\widetilde{#1}}

\newcommand{\Gr}{g_{\mathrm{R}}}
\newcommand{\GR}{G_{\mathrm{R}}}
\newcommand{\Cr}{c_{\mathrm{R}}}
\newcommand{\Dr}{\Delta_{\mathrm{R}}}
\newcommand{\Pt}{P_{\mathrm{T}}}
\newcommand{\PR}{P_{\Delta}}
\newcommand{\Leff}{L_{\mathrm{eff}}}
\newcommand{\R}{\mathrm{R}}
\newcommand{\Max}{\mathrm{Max}}
\newcommand{\Min}{\mathrm{Min}}
\newcommand{\der}{\mathrm{d}}
\newcommand{\ASE}{\mathrm{ASE}}

\newcommand{\OSNRtilde}{\mathrm{O\widetilde{SN}R}}
\newcommand{\OSNRhat}{\mathrm{O\widehat{SN}R}}

\newcommand{\Th}{\mathrm{th}}

\let\bigopsize\bigoplus
\def\bigoplus{{\scalerel*{\boldsymbol\oplus}{\bigopsize}}}

\newcommand{\appref}[1]{\hyperref[#1]{Appendix~\ref*{#1}}}

\makeatletter
\newcommand{\subalign}[1]{%
  \vcenter{%
    \Let@ \restore@math@cr \default@tag
    \baselineskip\fontdimen10 \scriptfont\tw@
    \advance\baselineskip\fontdimen12 \scriptfont\tw@
    \lineskip\thr@@\fontdimen8 \scriptfont\thr@@
    \lineskiplimit\lineskip
    \ialign{\hfil$\m@th\scriptstyle##$&$\m@th\scriptstyle{}##$\hfil\crcr
      #1\crcr
    }%
  }%
}
\makeatother

\newacronym{snr}{SNR}{signal-to-noise ratio}
\newacronym{osnr}{OSNR}{optical signal-to-noise ratio}
\newacronym{sinr}{SINR}{signal-to-interference-plus-noise ratio}
\newacronym{sdm}{SDM}{space-division multiplexing}
\newacronym{ssmf}{SSMF}{standard single-mode fiber}
\newacronym{mdg}{MDG}{mode-dependent gain}
\newacronym{mdl}{MDL}{mode-dependent loss}
\newacronym{csi}{CSI}{channel state information}
\newacronym{cdf}{CDF}{cumulative distribution function}
\newacronym{pdf}{PDF}{probability density function}
\newacronym{gue}{GUE}{Gaussian unitary ensemble}
\newacronym{wdm}{WDM}{wavelength-division multiplexed}
\newacronym{mmse}{MMSE}{minimum mean squared error}
\newacronym{mimo}{MIMO}{multiple-input multiple-output}
\newacronym{msle}{MSLE}{mean squared logarithmic error}
\newacronym{awgn}{AWGN}{additive white Gaussian noise}
\newacronym{rmse}{RMSE}{root-mean-squared error}
\newacronym{nli}{NLI}{non-linear interference}
\newacronym{gn}{GN}{Gaussian noise}
\newacronym{dge}{DGE}{dynamic gain equalizer}
\newacronym{gff}{GFF}{Gain flattening filter}

\newacronym{lp}{LP}{linearly-polarized}
\newacronym{qkd}{QKD}{quantum-key distribution}
\newacronym{ase}{ASE}{amplified spontaneous emission}
\newacronym{srs}{SRS}{stimulated Raman scattering}
\newacronym{isrs}{ISRS}{inter-channel stimulated Raman scattering}
\newacronym{sprs}{SpRS}{spontaneous Raman scattering}
\newacronym{fwm}{FWM}{four-wave-mixing}
\newacronym{imxt}{IMXT}{inter-mode cross-talk}

\begin{document}
\title{Closed-form Expression for the Power Profile in Wideband Systems with Inter-channel Stimulated Raman Scattering}
\author{Lucas~Alves~Zischler,~\IEEEmembership{Student~Member,~IEEE,}%
        ~Chiara~Lasagni,~\IEEEmembership{Member,~IEEE,}\\
        ~Paolo~Serena,~\IEEEmembership{Senior~Member,~IEEE,}%
        ~Alberto~Bononi,~\IEEEmembership{Senior~Member,~IEEE,}
        ~Giammarco~Di~Sciullo,~\IEEEmembership{Student~Member,~IEEE,}
        ~Divya~A.~Shaji,~\IEEEmembership{Student~Member,~IEEE,}
        ~Antonio~Mecozzi,~\IEEEmembership{Fellow,~IEEE,~Optica,}%
        ~and~Cristian~Antonelli,~\IEEEmembership{Senior~Member,~IEEE,~Fellow,~Optica}
\thanks{Manuscript received XXX xx, XXXX; revised XXXXX xx, XXXX; accepted XXXX XX, XXXX. This work was supported in part by the European Union under the Marie Sk\l odowska-Curie Grant Agreement No. 101120422 - Quantum Enhanced Optical Communication Network Security (QuNEST) and No. 101072409 - Optical Fiber Higher Order mode Technologies (HOMTech) (Corresponding Author: Lucas~Alves~Zischler)}%
\thanks{Lucas~Alves~Zischler, Giammarco~Di~Sciullo, Divya~A.~Shaji, Antonio~Mecozzi, and Cristian~Antonelli are with the Department of Physical and Chemical Sciences, University of L’Aquila, 67100 L’Aquila, Italy: (\mbox{e-mail: lucas.zischler@univaq.it}).}%
\thanks{Chiara~Lasagni, Paolo Serena, and Alberto Bononi are with the Department of Engineering and Architecture, Università degli Studi di Parma, 43124 Parma, Italy.}
}%

\markboth{}
{Closed-form Expression for the Power Profile in Wideband Systems with Inter-channel Stimulated Raman Scattering}

\maketitle

\begin{abstract}
  Wideband systems experience significant \gls*{isrs} and channel-dependent losses. Due to the non-uniform attenuation profile, the combined effects of ISRS and fiber loss can only be accurately estimated using numerical methods. In this work, we present an approximate closed-form expression for the channels' power profile accounting for these combined effects. We validate the proposed expression against numerical solutions in the case of CLU transmission, showing high accuracy for both single-span and multi-span fiber-optic links. Additionally, we derive an inverse expression, formulated as a function of the output power, which can be utilized to target a desired \gls*{osnr} profile through pre-emphasis of the launched channel powers.
\end{abstract}

\glsresetall


\glsresetall

\section{Introduction}

\IEEEPARstart{C}{apacity} requirements have been increasing exponentially, motivating the development of novel optical systems that operate beyond traditional bands~\cite{winzer2017scaling,hoshida2022ultrawideband}. Numerous experimental studies in the literature have explored transmission over increasingly wider bandwidths. In~\cite{fukuchi2002wideband}, the authors investigate SCL transmission and review the multiband technologies available at the time. In~\cite{renaudier2017first,renaudier2019107}, it is shown the transmission of $\sim$100~nm seamless spectra using novel semiconductor optical amplification. In~\cite{hamaoka2018150}, the authors optimize transmission capacities in a SCL setup by applying pre-emphasis to the launched power profile. In~\cite{galdino2020optical}, throughput is improved by employing different constellations across frequency channels. In~\cite{puttnam2022s}, the authors evaluate the benefit of utilizing distributed Raman amplification in a SCL transmission scenario.

Nevertheless, as the occupied bandwidth expands, non-linear effects and discrepancies in per-channel losses become increasingly significant~\cite{chraplyvy1984opticala,chraplyvy1990limitations}. Although \gls*{isrs} is also a concern in conventional C-band transmission, it becomes particularly significant in wideband systems, as the Raman efficiency increases with frequency separation, reaching a peak near a 14~THz difference~\cite{stolen1973raman}. The combined impact of \gls*{isrs} and frequency-dependent attenuation leads to noticeable distortions in the power spectral profile during propagation. This uneven power distribution can result in substantial disparities in \gls*{snr} values across channels and can result in capacity penalties if not properly mitigated~\cite{semrau2017achievable,cantono2018interplay,semrau2020modeling,okamoto2020study}.

By shaping the launch power spectrum, it is possible to pre-compensate for frequency-dependent distortions. This is achieved through a pre-emphasis profile, the estimation of which has been the subject of many works. In~\cite{lasagni2023generalized}, a closed-form expression is used to compensate for the \gls*{isrs}-induced tilt in the presence of \gls*{nli} for throughput maximization, but it does not address pre-compensation for channel-dependent losses. In~\cite{roberts2017channel,jarmolovivcius2024optimising}, iterative numerical methods are employed to minimize cost functions in order to achieve, respectively, a desired \gls*{snr} profile or throughput maximization under \gls*{isrs} and \gls*{nli}. These approaches rely either on closed-form solutions assuming constant attenuation or on numerical solutions of the power evolution equations. In~\cite{jiang2024optimization}, the authors employ a derivative-free optimization algorithm with closed-form \gls*{gn} model expressions to optimize throughput rates in a SCL setup. The methods proposed in~\cite{buglia2022impact,vasylchenkova2023launch,buglia2024throughput,yang2025122,jiang2025signal,jiang2024optimization} further incorporate distributed Raman amplification and use exhaustive iterative optimization techniques for throughput optimization.

To estimate capacity losses due to \gls*{isrs}-induced distortions, pre-emphasis profiles, and \glspl*{nli}, accurate estimation of how signal powers evolve along the fiber is required. In wideband systems, numerical solvers are usually employed to evaluate distortions in signal power profiles caused by \gls*{isrs} and frequency-dependent attenuation, since exact closed-form solutions are known only for the special case of constant attenuation and a linear Raman gain profile, as derived by Christodoulides~\cite{christodoulides1996evolution} and Zirngibl~\cite{zirngibl1998analytical}.

In this work, we provide an approximate closed-form solution for the signal power profiles accounting for the combined effects of \gls*{isrs} and frequency-dependent attenuation. Although an exact closed-form solution accounting for both \gls*{isrs} and frequency-dependent attenuation is not currently available in the literature, several approximate solutions have been proposed~\cite{poggiolini2018generalized,zefreh2020real,shevchenko2022maximizing,lasagni2023generalized}. In~\cite{poggiolini2018generalized,zefreh2020real}, a closed-form expression for the power profile evolution is derived to estimate \gls*{nli} levels. That approach relies on iteratively determining power profile coefficients, initially starting from a tentative answer in order to minimize a cost function. In~\cite{shevchenko2022maximizing}, the authors incorporate frequency-dependent attenuation into the signal power decay, while the \gls*{isrs}-induced gain or loss is computed from Zirngibl's solution. The work in~\cite{lasagni2023generalized} extends Zirngibl’s derivation to wider bandwidths by adopting a triangular approximation of the Raman gain profile. However, the derivation assumes a flat launch profile and constant attenuation, and the final expression replaces the constant attenuation with its frequency-dependent value.

The derivations presented in this work follow similar steps to those in Zirngibl's work, but retain the frequency dependence of the attenuation coefficient and assume a triangular approximation for the Raman gain profile, in order to derive an approximate closed-form expression for the channel power profiles. The proposed expression is based on the assumption that the total power decays exponentially along the fiber. The final solution depends on the launch power spectrum, the fiber attenuation profile, and the slope of the Raman gain profile. 

The closed-form expressions are validated against numerical solutions in different multiband transmission scenarios, with a special focus on the CLU case, where we demonstrate the model's accuracy under strongly frequency-dependent fiber loss. The solution is highly accurate when the transmission bandwidth falls within close range to the Raman triangular window, such as in CLU or SCL systems, but it becomes less accurate for scenarios with increasingly wider bandwidths. We further extend the solution to a multi-span scenario, where total power loss is compensated, but no in-line per-channel power equalization is applied, and show that the model accurately estimates channel powers even in the presence of accumulated frequency-dependent loss and \gls*{isrs}-induced distortion.

In addition, we provide an analytical solution to the inverse problem and demonstrate its use for pre-compensation of \gls*{isrs}-induced tilt, frequency-dependent losses, and band-dependent \gls*{ase} noise power. This is achieved by estimating the pre-emphasis profile that yields a desired \gls*{osnr} output spectrum through a simple iterative approach. The solution presented in this work specifically targets a desired \gls*{osnr} profile, rather than a \gls*{snr} profile, as nonlinearities are not considered in our analysis, limiting its accuracy, but significantly reducing computational complexity. The works in~\cite{roberts2017channel,jarmolovivcius2024optimising,jiang2024optimization,buglia2022impact,vasylchenkova2023launch,buglia2024throughput,yang2025122,jiang2025signal} optimize \gls*{snr} profiles using broad optimization methods, where the reduction in complexity lies on employing fast and accurate \gls*{gn} models for \gls*{nli} estimation. Even when executable in real time, such broad optimization algorithms remain computationally demanding. Simplifications can be achieved by focusing on \gls*{osnr}-limited transmission scenarios. Furthermore, several studies on optimal pre-emphasis estimation for wideband transmission have shown that \gls*{ase} noise is the dominant impairment~\cite{jiang2024optimization,buglia2022impact}, indicating that \gls*{osnr} often provides a tight upper bound to the actual \gls*{snr}. Therefore, the solution presented here can be used in combination with more sophisticated methods to narrow the search space of optimization algorithms, providing a rough estimate of the optimal profile that can further be fine tuned.

The remainder of this paper is organized as follows. In Section~\ref{sec:fundamental}, we review the fundamental equations describing \gls*{isrs} in the presence of frequency-dependent loss. Section~\ref{sec:closedform} presents the derivation of our closed-form expression for the transmitted signal powers as a function of propagation distance. In Section~\ref{sec:multispan}, we extend the solution to multi-span links. Section~\ref{sec:preemph} provides modified formulas to estimate a pre-emphasis profile that achieves a desired output \gls*{osnr} spectrum using the closed-form expressions. Section~\ref{sec:conclusion} is devoted to the conclusions.

\section{ISRS and Attenuation Profile Distortion on Power Spectra}
\label{sec:fundamental}

\begin{figure}[!t]
    \centering
    \includegraphics{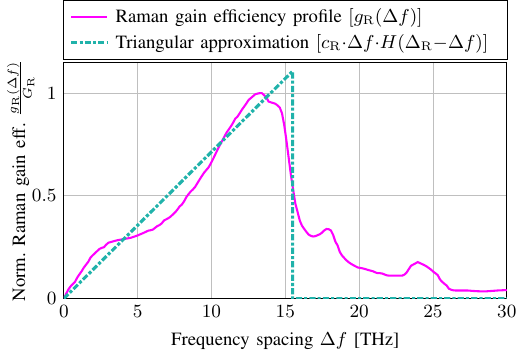}
    \caption{Normalized Raman gain efficiency profile based on~\cite{stolen1989raman,lin2006raman}, shown alongside its first-order and triangular approximations. The triangular model assumes the actual peak at $\Delta f=14$~THz~(${\Cr=\GR/14}$~THz) and spans a window of ${\Dr=15.5}$~THz. In the legend, $H(x)$ represents the Heaviside step function.}
    \label{fig:RamanGain}
\end{figure}

The power evolution of a signal propagating at frequency~$f_{i}$, in the presence of \gls*{isrs}, is known to obey the following equation~\cite[Eq.~(3)]{tariq2002computer}~\cite[Eqs.~(1)~and~(2)]{bromage2004raman}
\begin{equation}
  \begin{split}
    \frac{\der P_{i}(z)}{\der z}=&-\alpha(f_{i})P_{i}(z)+\sum_{f_{j}>f_{i}}\Gr(f_{j}-f_{i})P_{i}(z)P_{j}(z)\\
      &-\sum_{f_{j}<f_{i}}\frac{f_{i}}{f_{j}}\Gr(f_{i}-f_{j})P_{i}(z)P_{j}(z),
  \end{split}
  \label{eq:peqs}
\end{equation}
where $\alpha(f)$ is the frequency-dependent attenuation coefficient, and $\Gr(\Delta f)$ is the Raman gain efficiency profile, plotted in Fig.~\ref{fig:RamanGain}, with ${\GR = \max[\Gr(\Delta f)]}$ denoting the peak Raman gain efficiency. The Raman gain efficiency profile can be approximated by a triangular function with slope~$\Cr$ and window~$\Dr$, such that ${\Gr(\Delta f)\approx\Cr\cdot\Delta f\cdot H(\Dr-\Delta f)}$, where $H(x)$ is the Heaviside step function, as shown in Fig.~\ref{fig:RamanGain}. Within operational regimes of optical transmission, the contribution of \gls*{nli} to the total signal power is negligible. Therefore, although \gls*{nli} contribute to signal degradation, the accuracy of~\eqref{eq:peqs} is not reduced by its neglect.

The intensity of stimulated Raman scattering between two fields at two different frequencies is proportional to the overlap integral of the fields' lateral profiles, which is equal to the inverse of the effective mode area~\cite[Eq.~(33)]{rottwitt2003scaling}, making the Raman gain coefficient a bivariate function. This dependence is influenced by the fiber design~\cite{cordina2002changes}, and therefore the actual Raman gain response is highly specific to a given scenario. Nevertheless, experimental validations of closed-form \gls*{snr} expressions that neglect this frequency dependence have shown high accuracy in the SCL transmission window with E-band Raman pumps~\cite{yang2025122,buglia2023closed}, supporting the use of a frequency-independent approximation for the Raman gain slope. However, this dependence remains crucial in ultrawideband scenarios, where effective area values may diverge significantly across the transmission window~\cite{jiang2025signal}.

\begin{figure}[!t]
    \centering
    \includegraphics{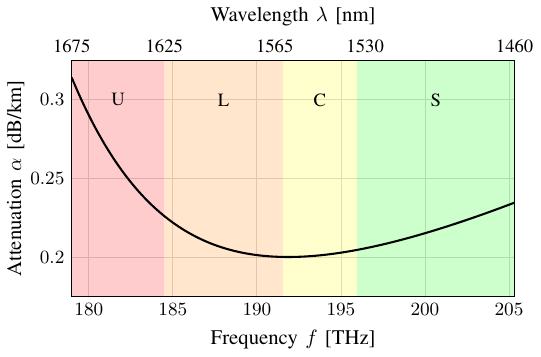}
    \caption{Attenuation profile for a doped silica fiber~\cite{walker1986rapid} and bands of transmission~\cite{hoshida2022ultrawideband}.}
    \label{fig:Attenuation}
\end{figure}

Losses in \glspl*{ssmf} typically exhibit a parabolic profile, with a global minimum located within the C-band. The attenuation profile assumed in this work is shown in Fig.~\ref{fig:Attenuation}.

Frequency-dependent losses are linear effects and can, in principle, be compensated in regions where non-linearities are negligible, since each channel can be treated independently. Nevertheless, the presence of \gls*{isrs} adds significant complexity into any compensation technique, as the power levels of different channels become correlated with each other due to Eq.~\eqref{eq:peqs}. Within conventional bands, assuming a constant attenuation profile, a closed-form solution for the \gls*{isrs}-induced tilt was derived in~\cite{christodoulides1996evolution,zirngibl1998analytical}. However, as seen in Fig.~\ref{fig:Attenuation}, this assumption becomes increasingly inaccurate as the bandwidth extends beyond the C-band, making the interplay between \gls*{isrs} and attenuation difficult to evaluate.

Disregarding frequency conversion losses (${f_{i}/f_{j}\approx 1}$), and assuming infinitely many equally spaced channels between $f_{\Min}$ and $f_{\Max}$ such that the per-channel power converges to the power spectral density, Eq.~\eqref{eq:peqs} becomes
\begin{equation}
  \frac{\partial S(f,z)}{\partial z}=-\alpha(f)S(f,z)+\Cr S(f,z)\hspace{-4pt}\int_{f-\Dr}^{f+\Dr}\hspace{-24pt}(f'-f)S(f',z)\der f'\hspace{-2pt},
  \label{eq:psdeqs}
\end{equation}
where $S(f,z)$ is non-zero only within the allocated spectrum, i.e., ${S(f,z) = 0}$ for $f \notin (f_{\Min}, f_{\Max})$. No exact closed-form solution is known for $S(f,z)$ for frequency-dependent $\alpha(f)$.

\section{Closed-Form Expressions for the Power Profiles}
\label{sec:closedform}

The main result of this work is the following expression for the spectral density profile, whose derivation is presented in~\appref{app:a}
\begin{equation}
  \begin{split}
    S(f\hspace{-1pt},\hspace{-1pt}z)\hspace{-3pt}&=\hspace{-3pt}S(f\hspace{-1pt},\hspace{-1pt}0)\exp\hspace{-2pt}\left[\hspace{-2pt}-\alpha(f)\hspace{-1pt}z\hspace{-2pt}+\hspace{-2pt}\Cr(\Gamma(f_{\R})\hspace{-2pt}-\hspace{-2pt}\Gamma(f))\hspace{-4pt}\int_{0}^{z}\hspace{-8pt}\Pt(z')\der z'\right]\hspace{-2pt},\\
  \end{split}
  \label{eq:snint}
  \raisetag{-8pt}
\end{equation}
where $\Pt(z)$ is the total power at position $z$, $\Gamma(f)$ is a shaping function accounting for the powers involved in the \gls*{isrs} process, and $f_{\R}$ is a reference frequency with no \gls*{isrs}-induced tilt. The shaping function is given by
\begin{equation}
  \Gamma\hspace{-1pt}(f)\hspace{-2pt}=\hspace{-2pt}\int_{f_{\Min}}^{f}\hspace{-10pt}\frac{\PR\hspace{-1pt}(f'\hspace{-4pt},\hspace{-0.5pt}0)-\Dr\left[S(f'\hspace{-4pt}+\hspace{-2pt}\Dr,\hspace{-0.5pt}0)\hspace{-3pt}-\hspace{-2pt}S(f\hspace{-4pt}-\hspace{-2pt}\Dr,\hspace{-0.5pt}0)\right]\hspace{-1pt}}{\Pt\hspace{-1pt}(0)}\der f'\hspace{-3pt},
  \label{eq:gamma}
\end{equation}
where $\PR(f,z)$ represents the total power within the $\pm\Dr$ Raman triangular window
\begin{equation}
  \PR(f,z)=\int_{f-\Dr}^{f+\Dr}S(f,z)df.
  \label{eq:prtri}
\end{equation}

In~\eqref{eq:gamma}, $\Gamma(f)$ is proportional to~\cite[Eq.~(18)]{lasagni2023generalized} for a flat launch profile. For bandwidths smaller than $\Dr$, $\Gamma(f)$ simplifies to
\begin{equation}
  \Gamma(f)=f-f_{\Min},~(f_{\Max}-f_{\Min})<\Dr.
\end{equation}

In our derivation leading to~\eqref{eq:snint}, the powers involved in the \gls*{isrs} process at frequency~$f$ are assumed to decay proportionally to the total power. However, this assumption becomes increasingly inaccurate as the bandwidth increases, since the powers within the \gls*{isrs} window account for an increasingly smaller fraction of the total power.

Integrating~\eqref{eq:psdeqs} over the entire spectrum nullifies the \gls*{isrs} contribution
\begin{equation}
  \int_{f_{\Min}}^{f_{\Max}}\frac{\partial S(f,z)}{\partial z}\der f=\frac{\der\Pt(z)}{\der z}=-\int_{f_{\Min}}^{f_{\Max}}\alpha(f)S(f,z)\der f,
\end{equation}
where $\Pt(z)$ is the total power within the entire occupied spectrum. Therefore, the instantaneous total power evolution is dictated solely by the attenuation and the power spectral profile, suggesting a weak contribution of \gls*{isrs} in the total power decay. Under constant attenuation, the total power evolution is dictated only by the constant attenuation coefficient, as shown in~\cite{zirngibl1998analytical}. However, since \gls*{isrs} shapes the power spectral distribution, it has an indirect effect on the total power longitudinal profile. Nevertheless, we assume an exponential decay of the total power, governed by a single effective attenuation coefficient $\alpha_{0}$, i.e., ${\Pt(z)\approx \Pt(0)e^{-\alpha_{0}z}}$. The total power integral in~\eqref{eq:snint} can then be approximated by
\begin{equation}
  \int_{0}^{z}\Pt(z')\der z'\approx \Pt(0)\frac{1-e^{-\alpha_{0}z}}{\alpha_{0}}.
  \label{eq:ptint}
\end{equation}

The total power attenuation coefficient can be obtained from (Derived in~\appref{app:b})
\begin{equation}
  \alpha_{0}\approx \sqrt[n]{\int_{f_{\Min}}^{f_{\Max}}\frac{\alpha^{n}(f')S(f',0)}{\Pt(0)}\der f'},
  \label{eq:alpha0}
\end{equation}
where $n\in\mathbb{N}^{+}$ is a free parameter, arising as a byproduct of the exponential approximation for the total power decay, as shown in~Appendix~B (see Eqs.~\eqref{eq:b2} and~\eqref{eq:b3}). The parameter $n$ provides an additional degree of freedom when applying the model, and can be universally optimized, as is discussed further in Sec.~III-A.

From the assumption of an exponential total power decay, we can also derive an expression for the shaping function at the zero-\gls*{isrs}-induced-tilt frequency\footnote{The value of $\Gamma(f_{\R})$ ensures energy conservation in \gls*{isrs} and is, strictly speaking, $z$-dependent. However, for simplicity, we evaluate it only at ${z=L}$ where the power profile is most relevant, and assume it to be constant over the fiber length. A more accurate power evolution profile over the entire span can be obtained by evaluating~\eqref{eq:fr} on the entirety of the $z$ domain, rather than just at ${z=L}$.} $\Gamma(f_{\R})$ as (Derived in~\appref{app:c})
\begin{equation}
  \begin{split}
    \Gamma(f_{\R})\approx&\frac{-1}{\Cr\Pt(0)\Leff}\ln\left[\int_{f_{\Min}}^{f_{\Max}}\frac{\alpha^{n}(f')S(f',0)}{\alpha_{0}^{n}\Pt(0)}\right.\\
    &\left. \quad\times e^{[\alpha_{0}-\alpha(f')]L-\Cr \Gamma(f')\Pt(0)\Leff}\der f'\right],
  \end{split}
  \label{eq:fr}
\end{equation}
where $\Leff=(1-e^{-\alpha_{0}L})/\alpha_{0}$ is the effective length of the total power profile, with $L$ denoting the fiber length. Replacing~\eqref{eq:ptint} in~\eqref{eq:snint} yields
\begin{equation}
  S(f,z)=S(f,0)e^{-\alpha(f)z+\Cr(\Gamma(f_{\R})-\Gamma(f))\Pt(0)\left(1-e^{-\alpha_{0}z}\right)/\alpha_{0}}.
  \label{eq:ssrs}
\end{equation}

Representing the signal as a finite sequence of CW tones, we arrive at a closed-form expression for the power profile of the propagated signals\footnote{An implementation of the closed-form equations shown in this work is available online at~\url{https://gitlab.com/lucaszischler/closed-form-isrs-wideband}.}
\begin{equation}
  P_{i}(z)=P_{i}(0)e^{-\alpha_{i}z+\Cr(\Gamma(f_{\R})-\Gamma(f_{i}))\Pt(0)\left(1-e^{-\alpha_{0}z}\right)/\alpha_{0}}.
  \label{eq:psrs}
\end{equation}

With finitely many discrete channels, the integrals in~\eqref{eq:alpha0} and~\eqref{eq:fr} are replaced by summations over the allocated channels, resulting in a complete closed-form solution. The shaping function $\Gamma(f)$ becomes
\begin{equation}
  \Gamma(f_{i})\hspace{-3pt}=\hspace{-3pt}\frac{\hspace{-1pt}\sum\limits_{j=0}^{i}\hspace{-0pt}B_{s}\hspace{-20pt}\sum\limits_{\hspace{8pt}|f_{k}\hspace{-1pt}-\hspace{-2pt}f_{i}|\hspace{-1pt}<\hspace{-1pt}\Dr}\hspace{-18pt}P_{k}(0)\hspace{-2pt}-\hspace{-2pt}\Dr\hspace{-3pt}\left[\hspace{-1pt}P_{\hspace{-2pt}j+\hspace{-1pt}\left\lfloor\hspace{-2pt}\frac{\Dr}{B_{s}}\hspace{-2pt}\right\rfloor}(0)+P_{\hspace{-2pt}j-\hspace{-1pt}\left\lceil\hspace{-2pt}\frac{\Dr}{B_{s}}\hspace{-2pt}\right\rceil}(0)\right]}{\Pt(0)}\hspace{-1pt},
\end{equation}
where $B_{s}$ is the frequency channel spacing.

\subsection{Numerical validation}
\label{ssec:numericalvalidation}

We start by evaluating the impact of the approximation order $n$ on the accuracy of the closed-form solution. To quantify the accuracy of the approximation, we define the total power error ratio at the fiber output as
\begin{equation}
  \epsilon_{P}=\frac{\sum_{i=1}^{N_{\text{ch}}}P_{i}(L)}{\sum_{i=1}^{N_{\text{ch}}}\widehat{P}_{i}(L)},
  \label{eq:epsilonP}
\end{equation}
where $N_{\text{ch}}$ is the channel count, $\widehat{P}_{i}(z)$ are the numerically computed power profiles from~\eqref{eq:peqs}, and $P_{i}(z)$ are the closed-form results from~\eqref{eq:psrs}. We evaluate $n$ over practical values for the Raman gain peak efficiency $\GR$, per-channel launch power $P_{i}(0)$, and fiber length $L$, across four bandwidth scenarios (C-band, CL, CLU, and SCLU). Channel spacing is assumed to be 50~GHz. Each variable is swept over 5 evenly spaced values within the following domains: ${\GR\in(0.3,0.4)}$~1/W/km, ${P_{i}(0)\in(-5,0)}$~dBm, and ${L\in(50,150)}$~km, for a total of 125 configurations for each bandwidth. The Raman gain slope coefficient $\Cr$ considers the actual peak at $14$~THz and triangular approximation with bandwidth $\Dr$ of $15.5$~THz. The actual Raman gain efficiency profile and triangular approximation shown in Fig.~\ref{fig:RamanGain}, based on data from~\cite{lin2006raman}. The considered attenuation profile shown in Fig.~\ref{fig:Attenuation}, derived from~\cite{walker1986rapid}. The power evolution equations are solved numerically using a 4$^{\Th}$ order Runge-Kutta method over 50 equal-length segments along the fiber span, with initial conditions defined by the channels launch powers.

\begin{figure}[!t]
    \centering
    \includegraphics{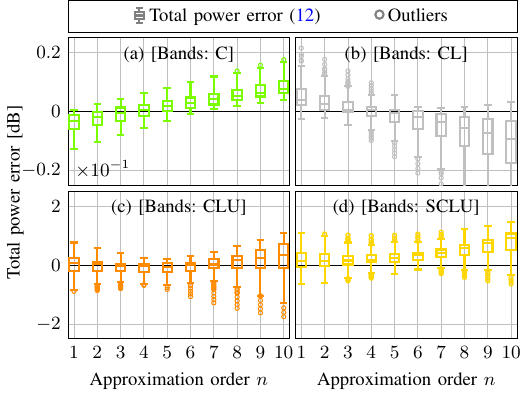}
    \caption{Box plots of the total power error ratio $\epsilon_{P}$ of the closed-form approximation, defined in~\eqref{eq:epsilonP}, as a function of the approximation order $n$. Each data point corresponds to a unique combination of $\GR$, $P_{i}(0)$, and $L$. The multiplicative factor shown in the lower-right corner of (a) indicates the scaling factor of the $y$-axis. Markers represent outlier values beyond 1.5 times the interquartile range.}
    \label{fig:OrderEr}
\end{figure}

The results are shown in Fig.~\ref{fig:OrderEr}, where each subplot refers to a specific bandwidth scenario. We see in Fig.~\ref{fig:OrderEr}(a) negligible error values for a C-band only scenario, regardless of the choice of $n$. In Figs.~\ref{fig:OrderEr}(b--d), approximation orders $n=3$ or $n=4$ produce the lowest average errors over all evaluated bandwidth sizes.

All subsequent simulations use the parameter values listed in Table~\ref{tab:parameters}, with Raman gain efficiency and attenuation profiles shown in Fig.~\ref{fig:RamanGain} and Fig.~\ref{fig:Attenuation}, respectively.

\begin{figure}[!t]
    \centering
    \includegraphics{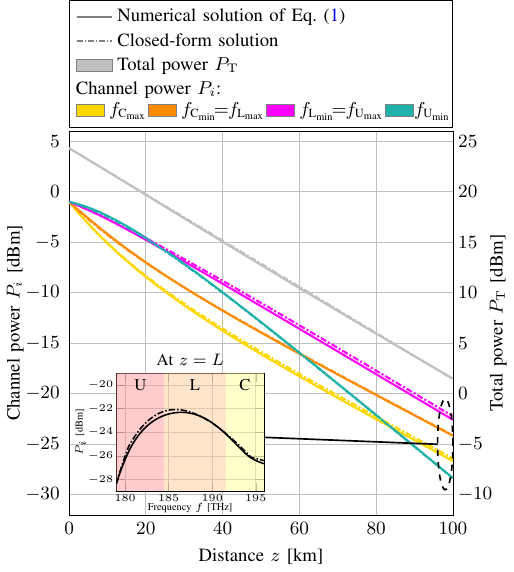}
    \caption{Longitudinal profiles of the total power and selected channels at the edges of the transmitted bands for a single-span CLU transmission, using the parameters in Table~\ref{tab:parameters}. Numerical solutions are shown as solid lines, while closed-form expressions are shown as dash-dotted lines. The closed-form solution is given by~\eqref{eq:psrs} for individual channel powers and by ${\Pt(z)=\Pt(0)e^{-\alpha_{0}z}}$ for the total power. The inset shows the power spectral profile at the fiber output (${z=L}$).}
    \label{fig:SingleSpan}
\end{figure}

Figure~\ref{fig:SingleSpan} compares the derived closed-form expressions with numerical solutions from~\eqref{eq:peqs} as a function of propagation distance for a CLU transmission scenario, where the fiber loss coefficient is strongly frequency-dependent. The total power longitudinal profile follows the exponential assumption closely. Colored lines correspond to channels at the edges of the transmitted bands, with the closed-form solution given by~\eqref{eq:psrs}. The inset of Fig.~\ref{fig:SingleSpan} compares the analytical and numerical power profiles at the fiber output. The excellent agreement between the two is self-evident. 

\begin{table}[!t]
    \centering
    \begin{tabular}{|c|c|cc|} 
         \hline
         \textbf{\makebox[1.14cm][c]{Spans}} & \textbf{\makebox[4.24cm][c]{Parameter}} & \multicolumn{2}{c|}{\textbf{Value}} \\\hline
         & Raman gain peak $\GR$ & \multicolumn{2}{c|}{0.4/W/km} \\
         & \parbox[c]{4cm}{\centering Raman gain eff. profile first-order coefficient $\Cr$ (14 THz)} & \multicolumn{2}{c|}{0.0286/W/km/THz} \\
         & Channel spacing $B_{s}$ & \multicolumn{2}{c|}{50 GHz} \\
         & Per-channel launch power $P_{i}(0)$ & \multicolumn{2}{c|}{-1 dBm} \\
         & Approximation order coefficient $n$ & \multicolumn{2}{c|}{3} \\\hline
         Single & Length $L$ & \multicolumn{2}{c|}{100 km} \\\hline
         & Span length $L_{s}$ & \multicolumn{2}{c|}{50 km} \\
         \multirow{-3}{*}{Multi} & Total length $L$ & \multicolumn{2}{c|}{250 km} \\\hline
         & Simulation sections per span & \multicolumn{2}{c|}{50} \\\hline
    \end{tabular}
    \vspace{0.1cm}

    \begin{tabular}{|c|c|c|c|c|c|}
        \hline
        \textbf{Bands} & \makebox[0.848cm][c]{C} & \makebox[0.848cm][c]{CL} & \makebox[0.848cm][c]{CLU} & \makebox[0.848cm][c]{SCL} & \makebox[0.848cm][c]{SCLU}\\\hline
        \hspace{-2pt}\textbf{Bandwidth [THz]}\hspace{-2pt} & \makebox[0cm][c]{4.05} & \makebox[0cm][c]{11.15} & \makebox[0cm][c]{16.65} & \makebox[0cm][c]{20.90} & \makebox[0cm][c]{26.05} \\\hline
        \textbf{Channel count} & \makebox[0cm][c]{81} & \makebox[0cm][c]{223} & \makebox[0cm][c]{333} & \makebox[0cm][c]{418} & \makebox[0cm][c]{521}\\\hline
    \end{tabular}
    \caption{Simulation parameters. Some parameters are specific to either single-span or multi-span scenarios.}
    \label{tab:parameters}
\end{table}

\begin{figure}[!t]
    \centering
    \includegraphics{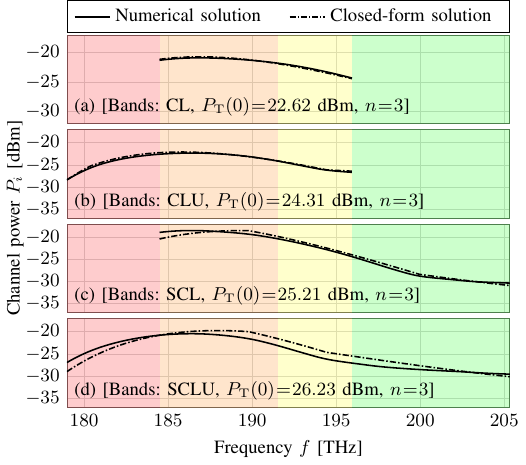}
    \caption{Power spectral profiles for increasingly wider bandwidths using the parameters in Table~\ref{tab:parameters}. Numerical solutions are shown as solid lines, while closed-form expressions are shown as dash-dotted lines. The corresponding total launch powers $\Pt(0)$ and approximation order coefficient~$n$ used are displayed in the figure.}
    \label{fig:Spectral}
\end{figure}

In Fig.~\ref{fig:Spectral}, the closed-form solution is evaluated for increasingly wider bandwidths, using the parameter values listed in Table~\ref{tab:parameters} for a single-span scenario. Figures~\ref{fig:Spectral}(a)~and~\ref{fig:Spectral}(b) indicate that the closed-form solutions are in excellent agreement with the numerical results when the considered spectrum does not greatly exceed the Raman triangular window. In Figs.~\ref{fig:Spectral}${\text{(c--d)}}$, the closed-form results underestimate the magnitude of the \gls*{isrs}-induced tilt. This discrepancy is possibly due to our approximation that the total power within the Raman triangular window is exponentially decaying, with equal rate as the total bandwidth power.

\section{Closed-form Expressions for Uncompensated Multi-Span Systems}
\label{sec:multispan}

The closed-form expression given in~\eqref{eq:psrs} describes the signal evolution along an unrepeated link. The \gls*{isrs}-induced tilt predominantly accumulates near the fiber input, where the total power $\Pt(z)$ is highest. As the total power attenuates along the fiber, the \gls*{isrs}-induced tilt increment vanishes. However, when signals are reamplified, the \gls*{isrs}-induced tilt increases again. With the combined effects of \gls*{isrs} and frequency-dependent losses, the power spectral profile can accumulate significant distortion over multiple uncompensated spans.

In-line amplifiers typically provide a constant gain over their operating bandwidth, with the gain value often being the sole adjustable parameter. An uneven gain profile can be equalized within the same amplification device with the aid of \glspl*{gff}. However, unless explicit information of the fiber link is provided to the amplifier, \glspl*{gff} do not compensate for channel-dependent losses. To mitigate accumulated distortions, long-haul links often employ \glspl*{dge} typically every 5$^{\Th}$ span to restore power levels~\cite{cantono2018interplay,cantono2020opportunities,lasagni2021modeling}. Even if desired, incorporating \glspl*{dge} in each in-line amplification device incurs in additional cost to the setup, so accumulated levels of \gls*{isrs}-induced distortion are expected in practical links.

In multi-band scenarios, different amplifiers are employed to cover distinct portions of the spectrum~\cite{ferrari2020assessment}. These amplifiers maintain total power levels within their respective windows, providing some degree of distortion compensation. Nevertheless, in-band distortions can still accumulate in a multi-span link.

Using~\eqref{eq:psrs}, we can also derive expressions for the signal profiles accounting for accumulated \gls*{isrs}-induced tilt and frequency-dependent losses over multiple spans. We consider a worst-case scenario where the amplification gain is uniform across the entire spectrum, even in a multi-band system, and is determined solely by the total power. Under this scenario, the total power at the beginning of any $k^{\Th}$ span remains constant (${\Pt^{(k)}(0)=\Pt^{(1)}(0)}$), and the gain of the $(k+1)^{\Th}$ in-line amplifier is equal to the total loss of the previous span
\begin{equation}
  G^{(k+1)}=\int_{f_{\Min}}^{f_{\Max}}\frac{S^{(k)}(f',L)}{\Pt^{(1)}(0)}\der f',
  \label{eq:spangain}
\end{equation}
where the output power spectral profile of each span can then be obtained from~\eqref{eq:ssrs} with the respective span parameters. The gain value in~\eqref{eq:spangain} is assumed to be constant over the evaluated spectrum. For a multi-band amplification scenario with constant per-band total power, the gain varies over frequency and can be trivially accounted for in our equations by replacing $G^{(k)}$ with its frequency-dependent value. Other schemes can be incorporated if the gain function is known. The gain is assumed to be lumped at the beginning of each span.

Assuming homogeneous spans, the power spectral density at the beginning of the $(k+1)^{\Th}$ span is given iteratively from the powers of the previous span by
\begin{equation}
  \begin{split}
    S^{(k+1)}\hspace{-2pt}(f,0)\hspace{-2pt}=&S^{(k)}\hspace{-2pt}(f,0)G^{(k+1)}\\
    &\times e^{-\hspace{-1pt}\alpha(f)\hspace{-1pt}L_{s}+\Cr\hspace{-1pt}[\Gamma^{(k)}(f^{(k)}_{\R}\hspace{-1pt})-\hspace{-1pt}\Gamma^{(k)}(f)]\Pt\hspace{-1pt}(0)\Leff^{(k)}},
  \end{split}
  \raisetag{12pt}
  \label{eq:s0pcte}
\end{equation}
where the effective length $\Leff^{(k)}$ and the shaping function $\Gamma^{(k)}(f)$ depend on the input spectral profile and the total power attenuation coefficient for the $k^{\Th}$ span. Given the launch power spectral density of the first span $S^{(1)}(f,0)$, we can iteratively compute the per-span values of $\alpha^{(k)}_{0}$, $\Gamma^{(k)}(f)$, $\Gamma^{(k)}(f^{(k)}_{\R})$, $G^{(k)}$, and $S^{(k)}(f,0)$ using~\eqref{eq:alpha0},~\eqref{eq:gamma},~\eqref{eq:fr},~\eqref{eq:spangain}, and~\eqref{eq:s0pcte}, respectively. The power profile at any distance along the link can then be obtained from~\eqref{eq:ssrs} with the corresponding span parameters.

\begin{figure*}[!t]
    \centering
    \includegraphics{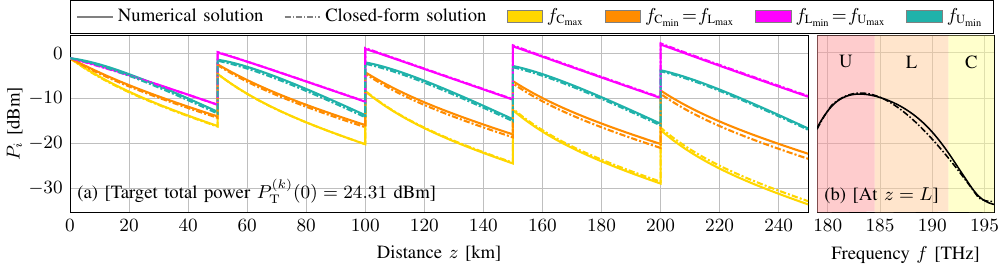}
    \caption{(a) Power profiles of selected channels at the edges of the transmitted bands, obtained via numerical methods, shown in solid lines, and closed-form expressions, shown in dash-dotted lines, considering the parameters listed in Table~\ref{tab:parameters} for a multi-span link and CLU transmission. (b) Power spectral profile at the link end (${z=L}$).}
    \label{fig:MultiSpan}
\end{figure*}

In Fig.~\ref{fig:MultiSpan}(a) and~\ref{fig:MultiSpan}(b), we compare numerical and analytical solutions for the longitudinal and spectral power profiles respectively, using the parameter values listed in Table~\ref{tab:parameters} for a CLU transmission setting. The numerical solution is obtained from~\eqref{eq:peqs}, solved via the Runge-Kutta 4$^{\Th}$ order method. At the end of each span (${z=k\cdot L_{s}}$), the power profiles are normalized so that the total power equals the initial launch power, thereby emulating the assumed amplification. The results show excellent agreement between numerical and analytical curves.

\section{Pre-Emphasis Estimation}
\label{sec:preemph}

The formulas provided so far require the values of the initial power profile in order to estimate the total attenuation coefficient $\alpha_{0}$, the shaping profile $\Gamma(f)$, and the shaping function at the zero-\gls*{isrs}-induced-tilt frequency $\Gamma(f_{\R})$. However, for pre-emphasis estimation, the output power spectral profile is set, and the input spectrum is to be determined. In what follows we derive the relevant expressions that are necessary to solve this problem.

As the shaping profile $\Gamma(f)$ is assumed to be $z$-independent, as discussed in~\appref{app:a}, it can be estimated from the output profile with
\begin{equation}
  \Gamma\hspace{-1pt}(f)\hspace{-2pt}=\hspace{-2pt}\int_{f_{\Min}}^{f}\hspace{-10pt}\frac{\PR\hspace{-1pt}(f'\hspace{-4pt},\hspace{-0.5pt}L)-\Dr\left[S(f'\hspace{-4pt}+\hspace{-2pt}\Dr,\hspace{-0.5pt}L)\hspace{-3pt}-\hspace{-2pt}S(f\hspace{-4pt}-\hspace{-2pt}\Dr,\hspace{-0.5pt}L)\right]\hspace{-1pt}}{\Pt\hspace{-1pt}(L)}\der f'\hspace{-3pt}.
  \label{eq:gammal}
\end{equation}

Following the previously discussed approximations, we can also obtain $\alpha_{0}$ and $\Gamma(f_{\R})$ as functions of the output powers, as given respectively by (Derived in~\appref{app:d})
\begin{equation}
  \alpha_{0}\approx \sqrt[n]{\int_{f_{\Min}}^{f_{\Max}}\frac{\alpha^{n}(f')S(f',L)}{\Pt(L)}\der f'},
  \label{eq:alpha0l}
\end{equation}
\begin{equation}
  \Gamma(f_{\R})\approx\int_{f_{\Min}}^{f_{\Max}}\frac{\Gamma(f')\alpha^{n}(f')S(f',L)}{\alpha^{n}_{0}\Pt(L)}\der f'.
  \label{eq:frl}
\end{equation}

With~\eqref{eq:alpha0l} and~\eqref{eq:frl}, we can obtain the input spectral profile that results in any arbitrary power profile at the fiber output
\begin{equation}
  S(f,0)=S(f,L)e^{\alpha(f)L-\Cr(\Gamma(f_{\R})-\Gamma(f))\Pt(L)\left(e^{\alpha_{0}L}-1\right)/\alpha_{0}}.
  \label{eq:premph}
\end{equation}

In some scenarios, setting the absolute values of the received powers for individual channels is not required, and only their relative powers are relevant. In such cases, we can replace the power spectral density profile in~\eqref{eq:alpha0l} and~\eqref{eq:frl} with a normalized profile, where only its shape is relevant, namely
\begin{equation}
  \begin{gathered}
    \widebar{S}(f,L)=\frac{S(f,L)}{\Pt(L)},\\
    \widebar{P}_{\Delta}(f,L)=\frac{\PR(f,L)}{\Pt(L)}.\\
  \end{gathered}
  \label{eq:snorm}
\end{equation}

With an added constraint on the total input power, we can obtain the actual output profile values. From the integration of~\eqref{eq:premph} over the entire band, by rewriting the total power integral as a function of the total input power as given by~\eqref{eq:ptint}, and after some manipulation, the total output power can be obtained in the following form
\begin{equation}
  \Pt(L)\hspace{-2pt}=\hspace{-2pt}\int_{f_{\Min}}^{f_{\Max}}\hspace{-6pt}\frac{\Pt(0)}{\widebar{S}(f',L)}e^{-\alpha(f')L+\Cr(\Gamma(f_{\R})-\Gamma(f'))\Pt(0)\Leff}\der f'\hspace{-3pt}.
\end{equation}

Multi-span pre-emphasis estimation can be performed with a similar approach to that of Section~\ref{sec:multispan}, where, for each span, we obtain a total power attenuation coefficient, reference frequency, and input power spectral profile. Under the constant total output power amplification scheme, the total per-span input power is an added constraint. In such cases, we can only estimate the pre-emphasis for a given normalized output power profile, as given in~\eqref{eq:snorm}. The $(k-1)^{\Th}$-span output power spectral profile is then iteratively obtained by
\begin{equation}
  S^{(k-1)}\hspace{-2pt}(f,\hspace{-2pt}L)\hspace{-2pt}=\hspace{-2pt}\frac{S^{(k)}\hspace{-2pt}(f,\hspace{-2pt}L)\hspace{-2pt}}{G^{(k)}}e^{\alpha(f)\hspace{-1pt}L_{s}\hspace{-1pt}-\Cr\hspace{-1pt}(\Gamma^{(k)}(f^{(k)}_{\R}\hspace{-1pt})-\hspace{-1pt}\Gamma^{(k)}(f))\Pt\hspace{-1pt}(0)\Leff^{(k)}}\hspace{-4pt}.
  \label{eq:sLpcte}
\end{equation}

The input power spectrum of any span can be obtained from~\eqref{eq:premph}, using the solutions from~\eqref{eq:sLpcte} and the respective span parameters.

\subsection{OSNR profile estimation}

The solutions in~\eqref{eq:premph} and~\eqref{eq:sLpcte} provide the power profiles. Nevertheless, it is often desirable to attain a given \gls*{osnr} profile. Disregarding non-linear effects, the closed-form solutions provided allow us to estimate a pre-emphasis profile for a desired \gls*{osnr} profile using a simple iterative approach.

With~\eqref{eq:sLpcte} and~\eqref{eq:spangain}, we are able to obtain the signal power profiles and per-span gains, respectively. In relevant transmission settings, the intensity of the amplification noise is negligible compared to the propagating signal, with insignificant impact on \gls*{isrs}. Therefore, we assume that its power evolves proportionally to the signal power, as given by
\begin{equation}
  S^{(k)}_{\ASE}(f,z)=S^{(k)}_{\ASE}(f,0)\frac{S^{(k)}(f,z)}{S^{(k)}(f,0)},
  \label{eq:ase}
\end{equation}
where $S^{(k)}_{\ASE}(f,z)$ is the accumulated \gls*{ase} noise power at the $k^{\Th}$ span. With the per-span gain values $G^{(k)}$, amplification noise figures, and the evolution profiles from~\eqref{eq:ase}, we are able to obtain the optical noise profiles at the link end ($z=L$). We assume that the noise powers are sufficiently low with respect to signal power values, such that they have no influence on the \gls*{isrs}-induced tilt.

As the total power constrain is at the beginning of each span in a multi-span scenario, due to in-line amplification, the actual output power values cannot be targeted. Nevertheless, normalized power values and \gls*{osnr} can be targeted, where the actual values are derived from the constrain condition. Given a desired normalized \gls*{osnr} profile $\OSNRtilde$, we can first assume that the normalized received powers $\widebar{S}(f,L)$ have the same shape ($\widebar{S}(f,L)\leftarrow\OSNRtilde$), and obtain an estimated normalized $\OSNRhat$ using the noise profiles from~\eqref{eq:ase}. We then update the received power profile with  
\begin{equation}
  \widebar{S}(f,L)\leftarrow\widebar{S}(f,L)\left(\frac{\OSNRtilde}{\OSNRhat}\right)^{\xi},
  \label{eq:iter}
\end{equation}
where $\xi$ is a step factor.

\begin{figure*}[!t]
    \centering
    \includegraphics{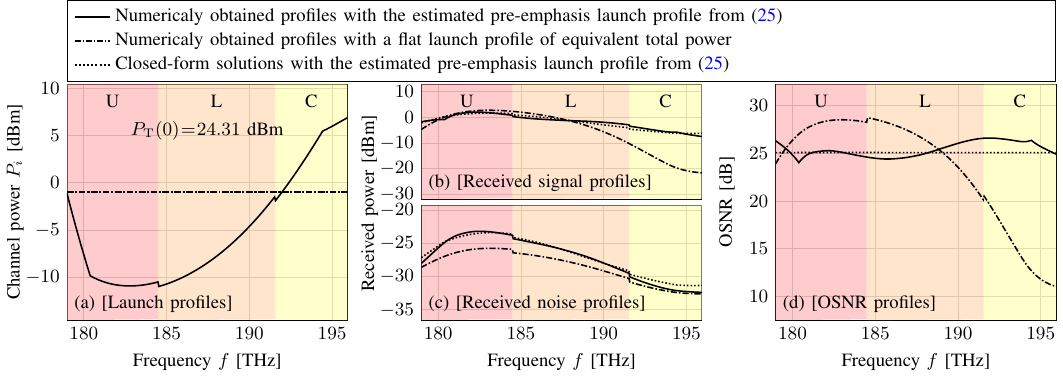}
    \caption{(a) Launch power profiles with equal total launch power. (b,~c) Received signal and accumulated \gls*{ase} noise profiles, respectively, after the boosting stage at the receiver. (d) \gls*{osnr} profiles for the considered launch profiles. In solid lines we have the numerically obtained profiles considering the estimated pre-emphasis targeting a flat \gls*{osnr} from~\eqref{eq:iter}. In dash-dotted lines we show, for comparison, the numerically obtained profiles considering a flat launch profile with equal power as the estimated pre-emphasis. The profiles derived from the closed-form solutions are presented alongside the numerical results in (b,~c,~d) with dotted lines.}
    \label{fig:PreEmphasis}
\end{figure*}

Figure~\ref{fig:PreEmphasis} shows the profiles for an estimated launch power pre-emphasis that targets a flat \gls*{osnr} over the CLU band. In the figure, solid and dash-dotted lines represent numerical solutions of~\eqref{eq:peqs} using, respectively, the estimated optimal pre-emphasis profile and a flat profile with equal total power. The dotted curves correspond to the closed-form solutions. Except for the per-channel launch powers, the simulation parameters are given in Table~\ref{tab:parameters} for a multi-span scenario. The amplifier noise figures are 5.5~dB for the C-band, 6.0~dB for the L-band, and 5.0~dB for the U-band~\cite{ferrari2020assessment,taengnoi2023ultra}. We disregard additive noise at the transmitter and assume a boosting stage at the receiver such that the total received power equals the total launch power. We iterate the normalized received profile using~\eqref{eq:iter} with a step coefficient of $\xi=1$, until the \gls*{rmse} between the desired and estimated normalized \gls*{osnr} over all frequency channels is smaller than~$10^{-5}$. We achieve this \gls*{rmse} value within 8 iterations.

Figure~\ref{fig:PreEmphasis}(a) shows that the estimated pre-emphasis results in slight offsets within the bands due to the different noise figure values. In Fig.~\ref{fig:PreEmphasis}(b) and (c), we see the received signal and noise powers, respectively. Without pre-emphasis, the received noise profile is overall smaller, but signal powers within the C-band are significantly lower due to \gls*{isrs} power transfer to the lower bands. The pre-emphasis profile compensates for this with higher launch powers within the C-band and reduced launch powers within the LU bands. In Fig.~\ref{fig:PreEmphasis}(d), we observe the resulting \gls*{osnr} profiles. With the pre-emphasis profile, we obtain a peak-to-peak \gls*{osnr} deviation of 2.58~dB, compared to a deviation of 17.73~dB with a flat launch profile.

\section{Conclusion}
\label{sec:conclusion}

In this paper, we derived a closed-form expression for the signal power profiles under the effects of~\gls*{isrs} and frequency-dependent attenuation. As both \gls*{isrs} and per-channel losses alter the power profile during propagation, the solution is non-trivial and relies on certain approximations. We assume that the effect of \gls*{isrs} on the total power is negligible and that the total power decays exponentially. The approximated closed-form expressions for both total and per-channel powers show strong agreement with numerical solutions.

From the single-span solution, we extend the expression for an uncompensated multi-span scenario, where distortions due to \gls*{isrs} and frequency-dependent attenuation accumulate at each span. The closed-form expression continues to show excellent agreement with numerical simulations.

We also present modified expressions for the approximated coefficients, reformulated as functions of the output powers. These expressions are then used to achieve a desired \gls*{osnr} profile at the receiver through an iterative approach. We validate the expressions in a CLU-band scenario for a multi-span link. The resulting output profiles demonstrate significant improvement in the shape of the received signal.

\appendices
\def\sectionautorefname{appendix}

\section{Derivation of Power Spectral Profile Given ISRS and Frequency-Dependent Attenuation}
\label{app:a}

In this section, we follow similar steps to those derived by Zirngibl~\cite{zirngibl1998analytical} and Lasagni~\cite{lasagni2023generalized}, with modifications to account for the frequency dependence of the attenuation coefficient throughout the derivation process. Dividing both sides of~\eqref{eq:psdeqs} by ${S(f,z)}$ and differentiating with respect to $f$
\begin{equation}
  \frac{\partial}{\partial f}\hspace{-2pt}\left[\frac{1}{S(f,z)}\frac{\partial S(f,z)}{\partial z}\right]\hspace{-2pt}=\hspace{-2pt}-\frac{\der \alpha(f)}{\der f}+\Cr\frac{\partial}{\partial f}\hspace{-4pt}\int_{f-\Dr}^{f+\Dr}\hspace{-26pt}(f'\hspace{-2pt}-\hspace{-2pt}f)S(f'\hspace{-2pt},z)\der f'\hspace{-2pt}.
  \label{eq:a1}
\end{equation}

We replace the Raman contribution term with ${-\Cr\beta(f,z)}$, where $\beta(f,z)$ is given by
\begin{equation}
  \begin{split}
    \beta(f,z)\hspace{-2pt}=\hspace{-2pt}\frac{\partial}{\partial f}\int_{f-\Dr}^{f+\Dr}\hspace{-26pt}(f\hspace{-2pt}-\hspace{-2pt}f')S(f'\hspace{-4pt},z)\der f'\hspace{-4pt}=&\PR\hspace{-2pt}(f,z)\hspace{-3pt}-\hspace{-3pt}\Dr\hspace{-2pt}\left[S(f\hspace{-3pt}+\hspace{-3pt}\Dr,z)\right. \\
    &\left. -S(f\hspace{-3pt}-\hspace{-3pt}\Dr,z)\right]\hspace{-2pt},
  \end{split}
\end{equation}
where $\PR(f,z)$ is the total power within the triangular approximation window, given in~\eqref{eq:prtri}.

The Raman contribution term is proportional to the power within the triangular window minus the power difference across the edges of the triangular window, multiplied by the triangular bandwidth. Assuming the tilt is small such that the power difference across the triangular bandwidth is negligible, and that the triangular bandwidth encompasses a sizeable portion of the total spectrum, we approximate the Raman contribution term as evolving proportionally to the total power:
\begin{equation}
  \beta(f,z)\approx\beta(f,0)\frac{\Pt(z)}{\Pt(0)}.
\end{equation}

Integrating~\eqref{eq:a1} with respect to $z$, from $0$ to $z$, we have
\begin{equation}
  \frac{\partial}{\partial f}\ln\left[\frac{S(f,z)}{S(f,0)}\right]=-z\frac{\der \alpha(f)}{\der f}-\Cr\frac{\beta(f,0)}{\Pt(0)}\int_{0}^{z}\Pt(z')\der z',
\end{equation}
and integrating with respect to $f$, from an arbitrary reference frequency $f_{\R}$
\begin{equation}
  \begin{split}
    \ln&\left[\frac{S(f,z)S(f_{\R},0)}{S(f,0)S(f_{\R},z)}\right]\hspace{-3pt}=\hspace{-3pt}\left[\hspace{-2pt}-\hspace{-1pt}\alpha(f')z\hspace{-2pt}-\hspace{-2pt}\Cr \Gamma(f')\hspace{-4pt}\int_{0}^{z}\hspace{-8pt}\Pt(z')\der z'\right]_{\hspace{-2pt}f'=f_{\R}}^{\hspace{-2pt}f}\\
    S&(f,z)\hspace{-2pt}=\hspace{-2pt}\left. \frac{S(f_{\R},z)}{S(f_{\R},0)}S(f,0)\exp\right\{\hspace{-2pt}-\left[\alpha(f)-\alpha(f_{\R})\right]z\\
    &\hspace{30pt}\left. +\Cr\left[\Gamma(f_{\R})-\Gamma(f)\right]\hspace{-4pt}\int_{0}^{z}\hspace{-8pt}\Pt(0)\der z'\right\},
  \end{split}
\end{equation}
where $\Gamma(f)$ is the result of an improper integration of the powers involved within the Raman contribution term, normalized by the total launch power, and given by~\eqref{eq:gamma}, where, arbitrary and with no impact to the final result, we set the lower integration boundary to the lower edge of the occupied spectrum.

We define $f_{\R}$ as the reference frequency with no \gls*{isrs}-induced tilt, given as
\begin{equation}
  S(f_{\R},z)=S(f_{\R},0)e^{-\alpha(f_{\R})z},
\end{equation}
where we then obtain~\eqref{eq:snint}.

\section{Derivation of the Total Power Attenuation Coefficient Expression}
\label{app:b}

As the \gls*{isrs} term in~\eqref{eq:snint} should not incur in any instantaneous gain or losses to the total power, integrating~\eqref{eq:psdeqs} over the entire spectrum, we have that the total power evolution is dictated by
\begin{equation}
  \frac{\der \Pt(z)}{\der z}=\int_{f_{\Min}}^{f_{\Max}}-\alpha(f')S(f',z)\der f',
  \label{eq:b1}
\end{equation}
where, under our approximation that ${\Pt(z)=\Pt(0)e^{-\alpha_{0}z}}$, we have that the total power power evolution can also be rewritten as
\begin{equation}
  \frac{\der \Pt(z)}{\der z}\approx-\alpha_{0}\Pt(0)e^{-\alpha_{0}z}.
  \label{eq:b2}
\end{equation}

From repeated derivations of~\eqref{eq:psdeqs} with respect to $z$, we also note that the \gls*{isrs} does not instantaneously contributes to any higher order derivatives of the total power
\begin{equation}
  \begin{split}
    \frac{\der^{n}\Pt(z)}{\der z^{n}}=&\int_{f_{\Min}}^{f_{\Max}}\frac{\partial^{n} S(f',z)}{\partial z^{n}}\der f'=\int_{f_{\Min}}^{f_{\Max}}\hspace{-16pt}[-\alpha(f)]^{n}S(f',z)\der f'\\
    \approx&(-\alpha_{0})^{n}\Pt(z),
  \end{split}
  \label{eq:b3}
\end{equation}
where we obtain infinitely many equations to solve for $\alpha_{0}$, resulting in a free parameter $n$. Isolating $\alpha_{0}$ from those solutions, and evaluation them at the fiber input (${z=0}$), we obtain~\eqref{eq:alpha0}.

\section{Derivation of the Shapping Function at the ISRS Tilt Reference Frequency}
\label{app:c}

Expanding $S(f,z)$ in~\eqref{eq:b3} with~\eqref{eq:snint}, considering the total power as exponentially decaying, and dividing both sides by ${(-1)^{n}}$ we have
\begin{equation}
  \begin{split}
    \alpha_{0}^{n}\Pt(z)\hspace{-2pt}=\hspace{-2pt}\int_{f_{\Min}}^{f_{\Max}}\hspace{-18pt}&\alpha(f')^{n}S(f',0)\\
    &\times e^{-\alpha(f')z+\Cr(\Gamma(f_{\R})-\Gamma(f'))\Pt(0)\frac{1-e^{-\alpha_{0}z}}{\alpha_{0}}}\der f'\hspace{-3pt}.
  \end{split}
  \label{eq:c1}
\end{equation}

Isolating $\Gamma(f_{\R})$ from~\eqref{eq:c1} we obtain a $z$-dependent expression for the shapping function at the zero-\gls*{isrs}-tilt frequency. As the \gls*{isrs} contribution of each channel evolves due to the evolving channel power spectral density, the reference zero-\gls*{isrs}-tilt frequency also changes. However, assuming a constant value for $f_{\R}$ does not results in significant discrepancies to the power profile over the $z$ domain, as seen in Fig.~\ref{fig:SingleSpan}, with the benefit of reduced complexity to the closed-form model. Therefore, evaluating the solution of $\Gamma(f_{\R})$ at the fiber end (${z=L}$), where it is of most interest, we obtain~\eqref{eq:fr}.

\section{Derivation of the Total Power Attenuation and ISRS Tilt Reference Frequency With the Power Profiles at the Fiber Output}
\label{app:d}

Considering an offset $x$ from the fiber output, we can rewrite~\eqref{eq:b3} as
\begin{equation}
  \frac{\der \Pt(L-x)}{\der z}\hspace{-2pt}=\hspace{-2pt}\int_{f_{\Min}}^{f_{\Max}}\hspace{-10pt}\alpha(f')^{n}S(f',L-x)\der f'\hspace{-2pt}=\hspace{-2pt}\alpha_{0}^{n}\Pt(L)e^{\alpha_{0}x}.
  \label{eq:d1}
\end{equation}

Isolating $\alpha_{0}$ from~\eqref{eq:d1}, and evaluation it at the fiber output (${x\rightarrow0}$), we obtain~\eqref{eq:alpha0l}.

The power spectral density term $S(f,L-x)$ at \eqref{eq:c1} can be expressed as a function of $S(f,L)$
\begin{equation}
  S(f,L-z)=S(f,L)e^{\alpha(f)x-\Cr(\Gamma(f_{\R})-\Gamma(f))\int_{L-x}^{L}\Pt(x')\der x'},
\end{equation}
where, under our assumption of an exponentially decaying total power, the total power integral can be given as a function of the received total power
\begin{equation}
  \int_{L-x}^{L}\Pt(z')\der z'\approx \Pt(L)\frac{e^{\alpha_{0}x}-1}{\alpha_{0}}.
\end{equation}

Under similar steps to~\appref{app:c}, by expressing ${S(f,L-x)}$ as a function of $S(f,L)$, and evaluating $f_{\R}$ at the fiber end, from the general arbitrary order solution given in~\eqref{eq:d1}, we obtain
\begin{equation}
  \begin{split}
    \Gamma(f_{\R})\hspace{-2pt}\approx\hspace{-2pt}\lim_{x\rightarrow 0}\frac{\alpha_{0}}{\Cr\Pt(L)(e^{\alpha_{0}x}-1)}\ln\left[\int_{f_{\Min}}^{f_{\Max}}\frac{\alpha^{n}(f')S(f',L)}{\alpha^{n}_{0}\Pt(L)}\right.&\\
    \left. \quad\times e^{[\alpha(f')-\alpha_{0}]x+\Cr \Gamma(f')\Pt(L)\frac{e^{\alpha_{0}x}-1}{\alpha_{0}}}\der f'\right].&
  \end{split}
\end{equation}
where the limit appears as the integral of the solution is indetermined at ${x=0}$. Solving for the limit we obtain~\eqref{eq:frl}.


\bibliographystyle{IEEEtran}
\bibliography{references}

\end{document}